\documentstyle[psfig]{mn}

\newif\ifAMStwofonts

\newcommand{\Mv}{\mbox{$M_{\mbox{\scriptsize V}}\,$}}

\newcommand{\pmag}{\mbox{$\stackrel{m}{\textstyle .}$}}

\newcommand{\degs}{\mbox{$^{o}$}}
\newcommand{\phour}{\mbox{$\stackrel{h}{\textstyle .}$}}

\newcommand{\arcsecs}{\mbox{$^{\prime\prime}$}}

\newcommand{\arcmins}{\mbox{$^{\prime}$}}

\newcommand{\Rsolar}{\mbox{$R_{\odot}\,$}}
\newcommand{\Msolar}{\mbox{$M_{\odot}\,$}}

\newcommand{\halpha}{\mbox{${\rm H}\alpha$}}

\newcommand{\kms}{\mbox{${\rm km\,s}^{-1}$}}

\def\goa{\mathrel{\mathchoice {\vcenter{\offinterlineskip\halign{\hfil
$\displaystyle##$\hfil\cr>\cr\sim\cr}}}
{\vcenter{\offinterlineskip\halign{\hfil$\textstyle##$\hfil\cr
>\cr\sim\cr}}}
{\vcenter{\offinterlineskip\halign{\hfil$\scriptstyle##$\hfil\cr
>\cr\sim\cr}}}
{\vcenter{\offinterlineskip\halign{\hfil$\scriptscriptstyle##$\hfil\cr
>\cr\sim\cr}}}}}

\ifoldfss
  \ifCUPmtlplainloaded \else
    \NewTextAlphabet{textbfit} {cmbxti10} {}
    \NewTextAlphabet{textbfss} {cmssbx10} {}
    \NewMathAlphabet{mathbfit} {cmbxti10} {} 
    \NewMathAlphabet{mathbfss} {cmssbx10} {} 
  \fi
  \ifAMStwofonts
    \ifCUPmtlplainloaded \else
      \NewSymbolFont{upmath} {eurm10}
      \NewSymbolFont{AMSa} {msam10}
      \NewMathSymbol{\upi}     {0}{upmath}{19}
      \NewMathSymbol{\umu}     {0}{upmath}{16}
      \NewMathSymbol{\upartial}{0}{upmath}{40}
      \NewMathSymbol{\leqslant}{3}{AMSa}{36}
      \NewMathSymbol{\geqslant}{3}{AMSa}{3E}

    \fi
  \fi
\fi 

\ifnfssone
  \newmathalphabet{\mathit}
  \addtoversion{normal}{\mathit}{cmr}{m}{it}
  \addtoversion{bold}{\mathit}{cmr}{bx}{it}
  \newmathalphabet{\mathbfit} 
  \addtoversion{normal}{\mathbfit}{cmr}{bx}{it}
  \addtoversion{bold}{\mathbfit}{cmr}{bx}{it}
  \newmathalphabet{\mathbfss} 
  \addtoversion{normal}{\mathbfss}{cmss}{bx}{n}
  \addtoversion{bold}{\mathbfss}{cmss}{bx}{n}
  \ifAMStwofonts
    \ifCUPmtlplainloaded \else
      \UseAMStwoboldmath
      \makeatletter
      \new@mathgroup\upmath@group
      \define@mathgroup\mv@normal\upmath@group{eur}{m}{n}
      \define@mathgroup\mv@bold\upmath@group{eur}{b}{n}
      \edef\UPM{\hexnumber\upmath@group}
      \new@mathgroup\amsa@group
      \define@mathgroup\mv@normal\amsa@group{msa}{m}{n}
      \define@mathgroup\mv@bold\amsa@group{msa}{m}{n}
      \edef\AMSa{\hexnumber\amsa@group}
      \makeatother
      \mathchardef\upi="0\UPM19
      \mathchardef\umu="0\UPM16
      \mathchardef\upartial="0\UPM40
      \mathchardef\leqslant="3\AMSa36
      \mathchardef\geqslant="3\AMSa3E
    \fi
  \fi
\fi 

\ifnfsstwo
  \DeclareMathAlphabet{\mathbfit}{OT1}{cmr}{bx}{it}
  \SetMathAlphabet\mathbfit{bold}{OT1}{cmr}{bx}{it}
  \DeclareMathAlphabet{\mathbfss}{OT1}{cmss}{bx}{n}
  \SetMathAlphabet\mathbfss{bold}{OT1}{cmss}{bx}{n}
  \ifAMStwofonts
    \ifCUPmtlplainloaded \else
      \DeclareSymbolFont{UPM}{U}{eur}{m}{n}
      \SetSymbolFont{UPM}{bold}{U}{eur}{b}{n}
      \DeclareSymbolFont{AMSa}{U}{msa}{m}{n}
      \DeclareMathSymbol{\upi}{0}{UPM}{"19}
      \DeclareMathSymbol{\umu}{0}{UPM}{"16}
      \DeclareMathSymbol{\upartial}{0}{UPM}{"40}
      \DeclareMathSymbol{\leqslant}{3}{AMSa}{"36}
      \DeclareMathSymbol{\geqslant}{3}{AMSa}{"3E}
    \fi
  \fi
\fi 

\ifCUPmtlplainloaded \else
  \ifAMStwofonts \else 
    \def\upi{\pi}
    \def\umu{\mu}
    \def\upartial{\partial}
  \fi
\fi

\title{The mass and radius of the M dwarf companion to GD\,448}
\author[P.F.L. Maxted et al.]
       {P.F.L. Maxted,$^1$
        T.R. Marsh,$^1$
        C. Moran,$^1$ 
        V.S. Dhillon,$^2$ 
        \newauthor
        and R.W.Hilditch $^3$ \\
        $^1$University of Southampton, Department of Physics \& Astronomy,
        Highfield, Southampton, S017 1BJ, UK\\
        $^2$Department of Physics and Astronomy, University of Sheffield, 
        Sheffield S3 7RH\\
        $^3$School of Physics and Astronomy, University of St. Andrews, North
Haugh, St Andrews, Fife, KY16 9SS, Scotland. }
\date{Accepted ??
      Received ??
      ??}

\pagerange{\pageref{firstpage}--\pageref{lastpage}}
\pubyear{1999}

\begin{document}

\maketitle

\label{firstpage}

\begin{abstract}
 We present spectroscopy and photometry of GD\,448, a  detached
white dwarf\,--\,M dwarf binary with a period of 2\phour47. We find that the
Na\,I\,8200\AA\ feature is composed of  narrow emission lines due to
irradiation of the M dwarf by the white dwarf within broad absorption lines
that are essentially unaffected by heating. Combined with an improved
spectroscopic orbit and gravitational red shift measurement from spectra of the
\halpha\ line, we are able to derive masses for the white dwarf and  M dwarf
directly ($0.41\pm0.01$\Msolar and $ 0.096\pm0.004$\Msolar, respectively). We
use a simple model of the Ca\,II emission lines to establish the radius of the
M dwarf assuming the emission from its surface  to be proportional to the
incident flux per unit area from the white dwarf. The radius derived is
0.125\,$\pm$\,0.020\,\Rsolar. The M dwarf appears to be a normal main-sequence
star in terms of its mass and radius and is less than half the size of its
Roche lobe. The thermal timescale of the M dwarf is much longer than the
cooling age of the white dwarf so we conclude that the M dwarf was
unaffected by the common-envelope phase. The anomalous width of the \halpha\
emission from the M dwarf remains to be explained, but the strengh of the line
may be due to X-ray heating of the M dwarf due to accretion onto the
white dwarf from the M dwarf wind.

 \end{abstract}


\section{Introduction}

 GD\,448 ($=$ WD\,0710$+$741, LP034$-$185) was shown by Marsh \& Duck (1996;
MD96 hereafter) to be a detached white dwarf\,--\,M dwarf binary with a period
of 2\phour47.  This is the shortest period known for such a binary and places
it in the centre of the ``period gap'' from 2 to 3 hours in which very few
cataclysmic variable stars (semi-detached  white dwarf\,--\,main-sequence
binaries) are found. From the cooling age of the white dwarf, MD96 found that
GD\,448 was born in the period gap and had never been a cataclysmic variable
star. They also found that the \halpha\  emission line from the heated face of
the M~dwarf is much broader than expected.  Since the thermal timescale of
the M~dwarf is much longer than the cooling-age of the white dwarf the radius
of the M~dwarf in GD\,448 is essentially unchanged from the radius at the end
of the common-envelope phase. This enables us to study the effect of the
common-envelope phase on the M~dwarf.  Unfortunately, the data of  MD96 were
of insufficient quality to establish the radius of the M~dwarf.

In this paper we present V and I band lightcurves of GD\,448 and much
improved spectroscopy which has enabled us to determine the radius of the
M~dwarf in GD\,448 and to improve the mass estimates for both components.

\section{Observations}

\subsection{Spectroscopy}
 We observed GD\,448 over four nights (11-14 January 1996) using the
double-beam spectrograph ISIS on the 4.2m William Herschel Telescope. Data
were obtained over three nights in variable conditions. We used the R1200R
grating in the  blue arm and the  R600R grating in the red-arm with  TEK CCD
detectors to obtain spectra covering 400\AA\ around \halpha\  at a dispersion
of 0.4\AA\ per pixel  and an 800\AA\ region covering the Na\,I\,8200\AA\
doublet and Ca\,II infrared triplet at a dispersion of 0.8\AA\ per pixel. A
total of 161 spectra were obtained in each spectral region. A 0.9\arcsec\ wide
slit which projected onto 1.6 pixels at the detectors was used to ensure
accuracy in the measured radial velocities. Exposure times were
300s\,--\,500s. Optimal extraction was used to extract spectra from our CCD
images. All the observation were interspersed with  observations of a
copper-argon arc at least every 40 minutes and before and after every
re-pointing of the telescope. Spectra of the arc were extracted at the same
detector position as the object spectra. The wavelength scale determined from
a polynomial fit to measured arc line positions was interpolated in time for
the object spectra.

 The spectra were flux calibrated using spectra of BD$+26$\,2606 (Oke 1983) to
correct for the wavelength-dependent instrument sensitivity.  No correction
for slit losses was attempted so the spectra have an arbitrary absolute flux
scale. Telluric features in the red-arm data were removed using the technique
of Wade \& Horne (1988). The rapidly rotating B-type star HD\,84937 was used to
determine the telluric absorption spectrum. The removal is naturally imperfect
as the strength of the telluric absorption  had to be estimated from the
airmass at which each spectrum of GD\,448 was observed. 

\subsection{Photometry}
 We used the  1m Jacobus Kapteyn telescope to obtain V and I band CCD images
of GD\,448. The detector used was a  TEK CCD with $1024^2$ pixels giving an
image scale of 0.34\arcsec\ per pixel. Observations were obtained during the
first 1-2 hours of the nights 28  March to 2 April 1996. A total of 84 useful
images were obtained with the I filter and 30 in the V filter. Normalized
average images of the twilight sky were used to apply flat-field corrections to
all the images. 

%
%
\begin{table}
  \caption{\label{CoordTable} Position of GD448 and comparison stars
 Coordinates are equinox 2000.0,
epoch 1996.24 and are accurate to a few arcseconds. }
  \begin{tabular}{@{}lrr@{}}
Star & \multicolumn{1}{c}{RA}& \multicolumn{1}{c}{Dec}\\
GD\,448  &$07^h 17^m 09.6^s $&$+74\degs 00\arcmins 42\arcsecs $ \\
C1       &$07^h 17^m 03.9^s $&$+74\degs 03\arcmins 32\arcsecs $ \\
C2       &$07^h 16^m 41.3^s $&$+74\degs 00\arcmins 23\arcsecs $ \\
\end{tabular}
\end{table}

%
%
\begin{figure}
\caption{\label{LightcurveFig}
The  V and I lightcurves of GD\,448 (upper and lower panels, resp.). }
\psfig{file=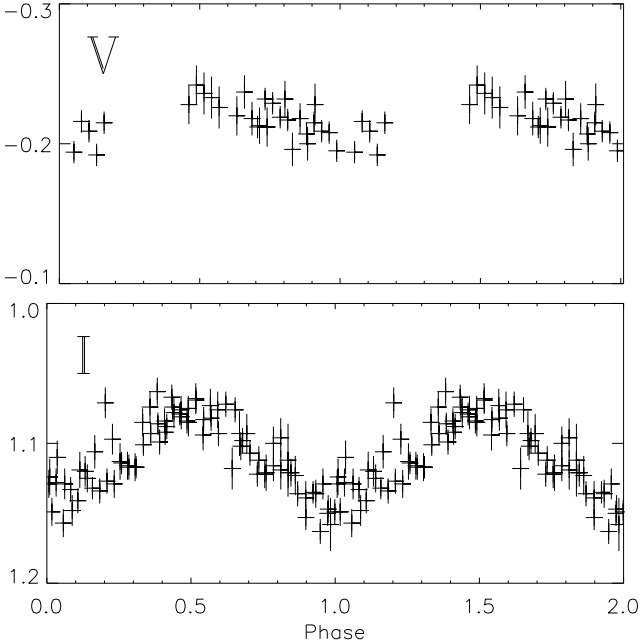}
\end{figure} 

 We used profile fitting to determine differential magnitudes of 8 stars in
the images including GD\,448, all of which were used to determine the shape of
the profile to be used in each image. After careful inspection of the images
and the resulting magnitudes of these stars we chose the comparison star C1
and a check star C2 ($\equiv$ GSC\,4372\,00331) whose positions are given in
table~\ref{CoordTable}. The positions of these  stars and GD\,448  were
measured from our images using positions for 5 Hubble Guide Star Catalogue
stars to establish the linear astrometric solution. The magnitude difference
between these stars is seen to be constant to within $\approx 0\pmag01$. The
lightcurve of GD\,448 with respect to C1 is shown in Fig.~\ref{LightcurveFig}
where the phase has been calculated using the ephemeris given below. The I
band lightcurve clearly shows the effect of the irradiation of the M dwarf by
the white dwarf. The incomplete coverage of the V lightcurve makes it
difficult to judge whether the variabilty at this wavelength is due to
irradiation, although a cosine function gives a good fit to the incomplete
lightcurve and suggests the amplitude of the V lightcurve is half that of the
I lightcurve. 

\section{Results}

\subsection{The Ca\,II triplet} 

%
%
\begin{figure}
\caption{\label{RVFig}
The measured radial velocities for the Ca\,II triplet emission
 lines (upper panel) and adopted fit. }
\psfig{file=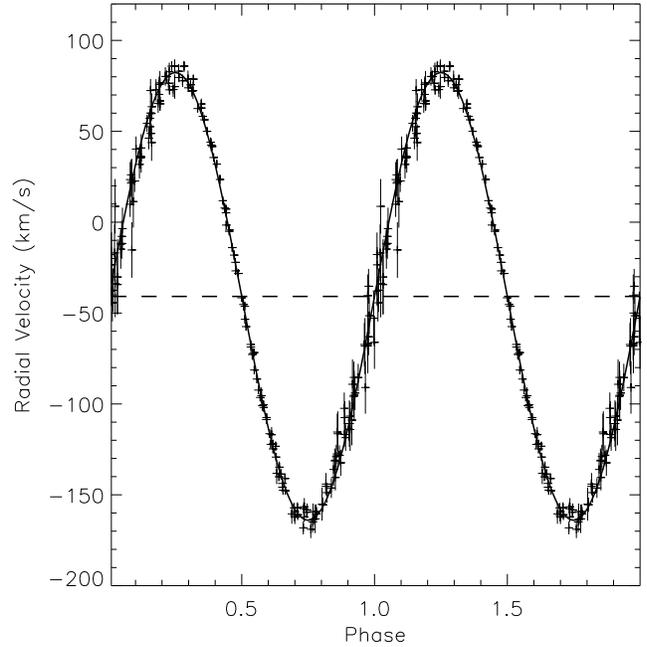,width=85mm}
\end{figure}

%
%
\begin{figure*}
\caption{\label{TrailFig}
 Trailed and phase-binned spectra of the Ca\,II\,8542\AA\ line (left panel), the
model spectra for the simple model of Ca\,II emission  and the
adopted value  $f$ (middle panel) and the \halpha\ emission line after
subtraction of the white dwarf absorption (right panel). 
}
\psfig{file=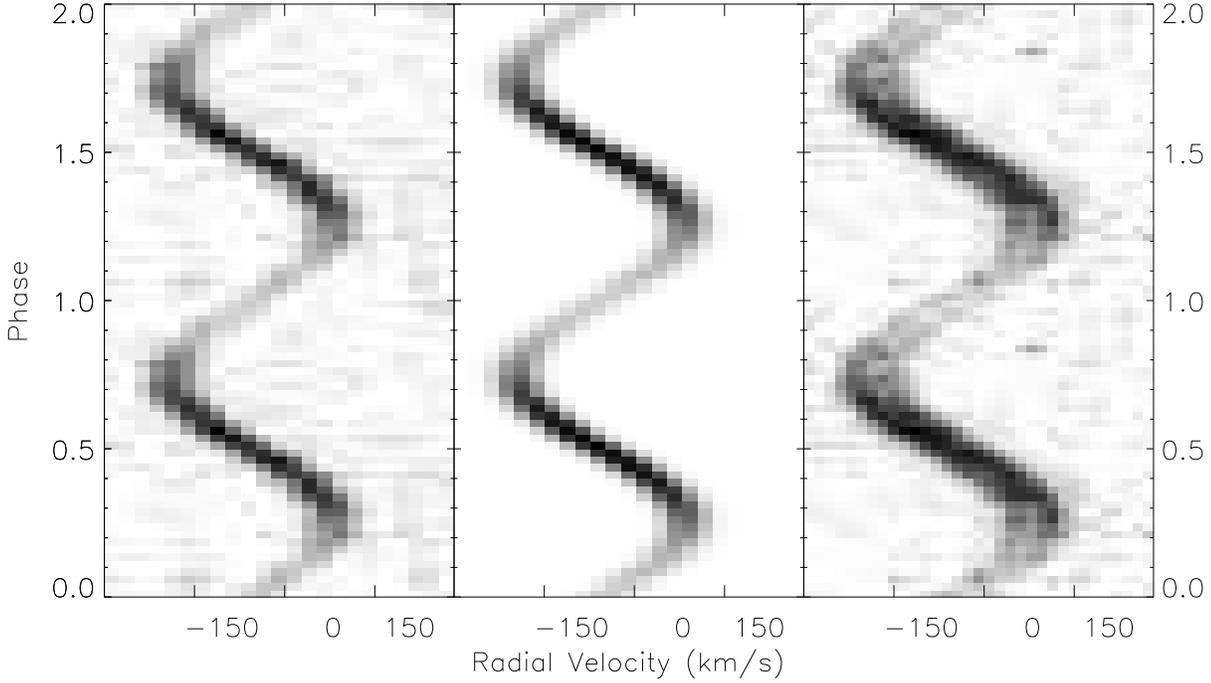}
\end{figure*}

 The red-arm spectrum shows the Ca\,II triplet emission lines seen by MD96.
These are of roughly equal strength and show sinusoidal variations of
intensity and Doppler shift with phase. This can be seen in Figs.~\ref{RVFig},
\ref{TrailFig} and \ref{EWFig}, which are described in more detail below.

%
%
\begin{table}
  \caption{\label{CaIIParTable} The parameters of the Gaussian profiles used 
to fit the  Ca\,II emission lines.}
  \begin{tabular}{@{}lrl@{}}
parameter   & Value                       & Notes \\
FWHM(\kms)  & $58.6 \pm 0.5 $ & Mean width of lines. \\
            &                 & No phase variation seen.   \\
h(8662)/h(8542) & $0.830 \pm 0.010$   & Ratio of heights \\
h(8498)/h(8542) & $0.708 \pm 0.007$   & Ratio of heights \\
 $A^1$    & $0.409 \pm 0.002$       & \\ 
 $B^1$    & $-0.267 \pm 0.003$      & \\
\multicolumn{3}{l}{1: Fit to height dependence on phase $\phi$,
 $h = A + B\cos(\phi)$}\\ 
\end{tabular}
\end{table}

 We used a series of Gaussian fits to individual spectra to measure the radial
velocity  of the Ca\,II triplet in a similar manner to MD96. The spectra are
normalised to a continuum value near the Ca\,II lines determined from a
cosine fit to the I-band lightcurve. We find $C_I = 1 - 0.03\cos(\phi)$, where
$C_I $ is the continuum value and $\phi$ is the orbital phase in radians,
phase zero being times at which the M-star is closest to the Earth. The ratio
of the heights of the Ca\,II lines was held fixed at a value determined from a
previous fit  and the variation of the height of the lines with phase was set
by a fit to the measured individual heights of the form $A + B \cos(\phi)$. We
detected no variation of the width of the lines with phase so this height
variation has the same form as the equivalent width variation seen in
Fig.~\ref{EWFig}. Values of the parameters used in the fit are given in
table~\ref{CaIIParTable}. The reduced chi-squared values for the fits were
typically 0.9\,--\,1.1. 

%
%
\begin{table}
\caption{\label{RVFitTable} Circular orbit fits}
\begin{tabular}{@{}lrr@{}}
Spectral line(s) &$ \gamma(\kms)$ & $K(\kms)$  \\
Ca\,II IR triplet emission & $-40.8 \pm 0.35$& $123.1 \pm 0.4$ \\
\halpha\ absorption        & $-24.2 \pm 0.6 $& $-33.5 \pm 0.9$ \\
\halpha\ emission          & $-38.1 \pm 0.4 $& $125.4 \pm 0.6$ \\
Na\,I\,8200\AA\ absorption   & $-38.6 \pm 1.5 $& $142 \pm 3$     \\
Na\,I\,8200\AA\ emission     & \multicolumn{1}{c}{''}& $120 \pm 3$      \\
8800.3\AA\   absorption    &  \multicolumn{1}{c}{---}& $148 \pm 5$ \\
\end{tabular}
\end{table}

We used a periodgram of  our radial velocity measurements combined with those
of MD96 to confirm the orbital period, $P$. The cycle count between the two
data sets is confirmed by the lightcurve. From the sine curve fit of the form
$\gamma_{\rm M} + K^{\rm Ca\,II}_{\rm M}\sin(\phi)$ to the combined radial
velocity data shown in Fig.~\ref{RVFig} we find the following ephemeris:
\[{\rm HJD}(T_0) = 2450094.69006(7) + 0.10306437(7)E,\] where figures in
parentheses are 1-$\sigma$ uncertainties in the final digit and $T_0$ refers
to dates at which the M-star is closest to the Earth. The values of
$\gamma_{\rm M}$ and $K^{\rm Ca\,II}_{\rm M}$ determined from this fit are given
in table~\ref{RVFitTable}. This ephemeris agrees with that of MD96 once a
superfluous digit in their quoted period is accounted for.

%
%
\begin{figure*}
\caption{\label{RedArmFig}
The average red-arm spectrum after removal of the Ca\,II radial
velocity Doppler shift.}
\psfig{file=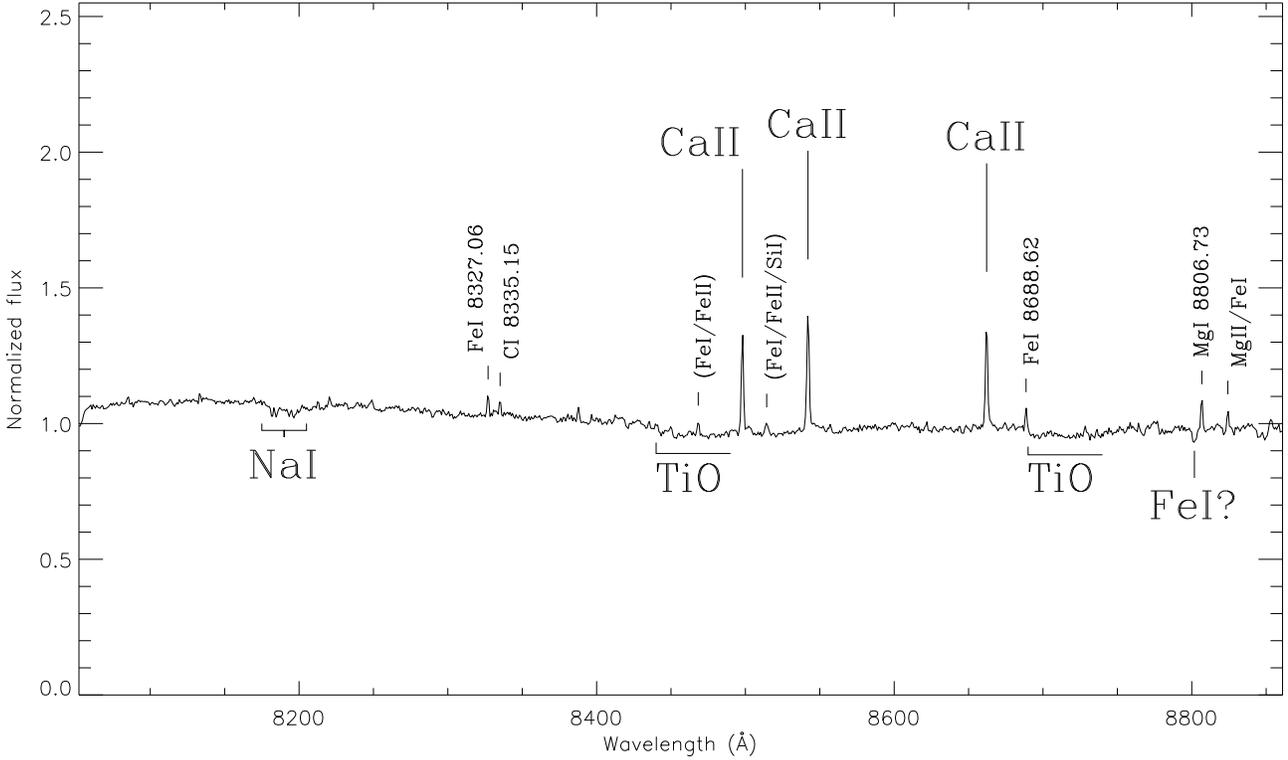}
\end{figure*}  

The average red-arm spectrum after subtraction of the Doppler shift due to
this orbit is shown in Fig.~\ref{RedArmFig}. In addition to the Ca\,II triplet,
there are several other emission lines visible. Some of these are labelled
with tentative identifications which are based on a simple model of emission
from an optically thin gas with a range of effective temperature around
{\mbox 5000\,-\,10\,000K}. Also visible is the Na\,I\,8200\AA\ feature, which
is distorted by the averaging process, as well TiO and other absorption
features.  

\subsection{The Na\,I doublet.}

%
%
\begin{table*}
  \caption{\label{NaIParTable} The parameters of the Gaussian profiles used 
to fit the  Na\,I doublet. Parameters in bold type are fixed quantities.}
  \begin{tabular}{@{}lrrrrr@{}}
Wavelength(\AA)&\bf{8183.3}&\bf{8194.8}&\bf{8183.3}&\bf{8194.8}&$8207.5\pm0.4$\\
FWHM(\AA)$^1$&$3.3\pm0.2$&$3.8\pm0.2$&\multicolumn{2}{c}{$1.8\pm0.1$}&$2.5\pm0.9$\\
Height&$-0.079\pm0.004$&$-0.073\pm0.004$&$1.07\pm0.09$&$0.99\pm0.08$&
       $-0.010\pm0.002$\\
$A    $  & --- & --- & $0.066\pm0.003$&$0.057\pm0.003$& --- \\
$B    $  & --- & --- & $-0.038\pm0.004$&$-0.048\pm0.004$& --- \\
\end{tabular}
\end{table*}

%
%
\begin{figure}
\caption{\label{NaITrailFig}
Trailed and phase-binned spectra around the Na\,I 8200\AA\ absorption
line (left panel) and the best fit using multiple Gaussian fits (right
panel).
}
\psfig{file=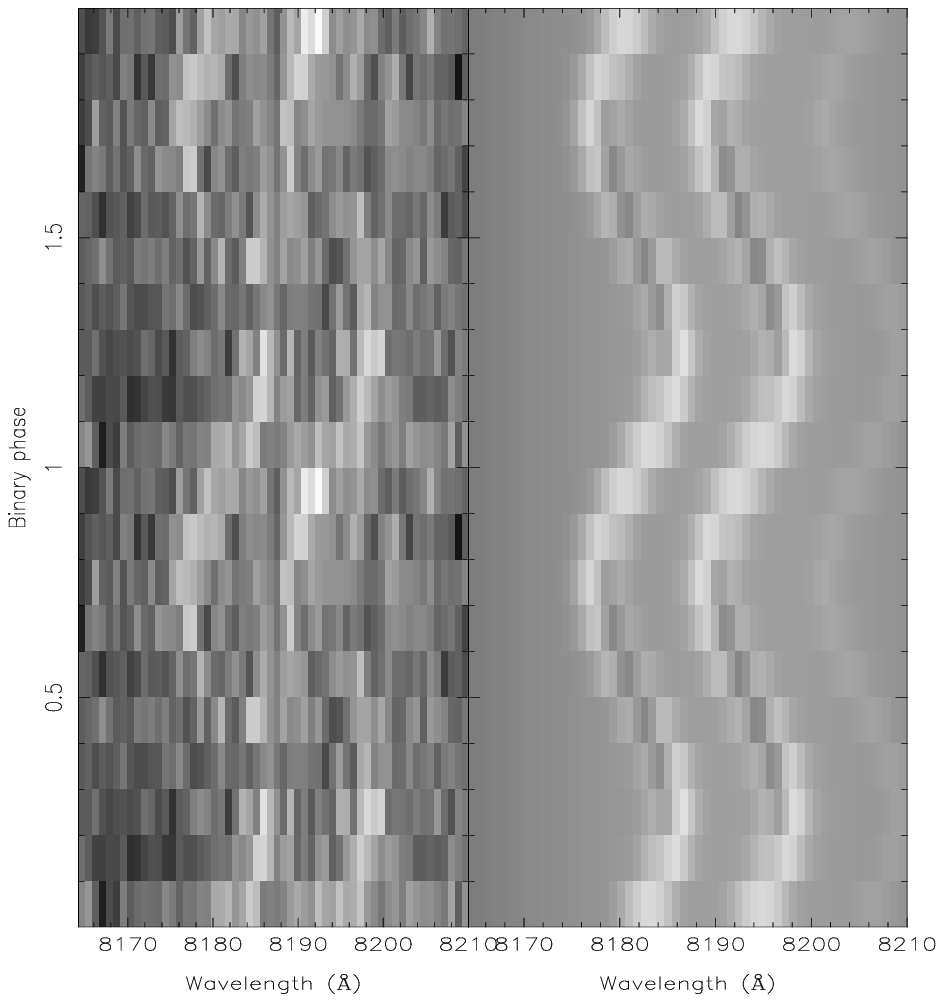,width=8.5cm}
\end{figure}

 We experimented with several models to account for the variation with phase
of the Na\,I doublet. The most successful and easiest to interpret is that of
narrow emission lines, similar to the Ca\,II emission lines, with
semi-amplitude $K^{\rm Na\,I}_{\rm M}$ superimposed on  broad absorption lines
which are constant with phase, i.e., with  the same semi-amplitude as the
centre-of-mass of the M dwarf, $K_{\rm M}$. To measure the values of $K^{\rm
Na\,I}_{\rm M}$ and $K_{\rm M}$ we used a fitting method similar to Moran
et~al. (1997), in which a simultaneous Gaussian fit to all the spectra is
used to determine the orbital parameters.   A single Gaussian was used to
model each emission and  absorption component for both  Na\,I lines and a
third Gaussian was used to account for weak absorption line nearby. The
parameters of these Gaussians are given in table~\ref{NaIParTable}. The
variation in height of the emission lines was established from a cosine fit to
measured individual heights. note that this is similar to the variation
seen in the Ca\,II emission lines. The  reduced chi-squared value of the final
fit is 1.05 (9285 data points) which is acceptable given the difficulties in
removing telluric features in this region. The observed Na\,I line spectra and
fits have been coadded into phase bins to produce the trailed spectra  shown
in Fig.~\ref{NaITrailFig} .

 We were able to confirm the value of $K_{\rm M}$ using the absorption line
seen near 8800\AA\ in Fig.~\ref{RedArmFig}, which we tentatively identify as
FeI\,8801.8\AA. We used a fit of a single Gaussian to this line to determine
$K_{\rm M}$. The line shows no variation of depth or width with phase and is
unaffected by telluric features. The nearby MgI emission line was accounted
for in the fit and was found to behave in a similar way to the Ca\,II lines. We
find $K_{\rm M} = 148 \pm 5\kms$, in good agreement with the value determined
from Na\,I. The weighted average of the two values is $144\pm3\kms$, which is
our adopted value for $K_{\rm M}$. 

\subsection{The \halpha\ line.}
%
%
\begin{table}
\caption{The parameters of the Gaussian profiles used to fit \halpha.
}
\label{HaParTable}
\begin{tabular}{lcc}
\multicolumn{1}{l}{Component}& \multicolumn{1}{l}{FWHM(\AA)}  & \multicolumn{1}{l}{Height} \\
Emission line           & $2.08 \pm0.02 $ & $ 1.030\pm0.0006^a $\\  
Absorption component 1  & $1.321\pm0.034$ & $-0.175\pm0.002$  \\
Absorption component 2  & $6.87 \pm0.30 $ & $-0.070\pm0.002$  \\
Absorption component 3  & $28.36\pm0.90 $ & $-0.109\pm0.003$  \\
Absorption component 4  & $67.35\pm2.41 $ & $-0.117\pm0.003$  \\
Absorption component 5  & $153.6\pm4.40 $ & $-0.074\pm0.003$  \\
\multicolumn{3}{l}
 {$a$: The actual height of the emission is given by this value } \\
\multicolumn{3}{l}
 {~~~  multiplied by the cosine variation described in the text.}   \\
\end{tabular}
\end{table}

 The contribution from the M dwarf to the continuum near \halpha\ is
negligible so the continuum was set to unity using a linear fit to the
continuum 8000-8500\kms either side of the line. We used five Gaussian
profiles to model the absorption profile and a single Gaussian profile to
measure the emission feature. A series of fits were used to establish the
height variation of the emission line, which is similar to the Ca\,II emission.
We also established the widths and heights of the Gaussian profiles given in
table~\ref{HaParTable}, and the apparent systemic radial velocity,
$\gamma_{\rm M}$ and semi-amplitude, $K^{\rm \halpha}_{\rm M}$ of the \halpha\
emission line given in table~\ref{RVFitTable} using a simultaneous fit to all
the spectra. 

 We were concerned that the variation of the \halpha\ emission line might
affect our estimates of $\gamma_{\rm WD}$ and $K_{\rm WD}$ --  the apparent
systemic radial velocity and semi-amplitude of the white dwarf. To avoid this
we excluded a region on either side of the predicted emission line position
from the fitting process with similar width to the FWHM of the emission line
($\sim 100\kms$). We then used a simultaneous fit to all the spectra to
measure the $\gamma_{\rm WD}$ and $K_{\rm WD}$. The results vary slightly
depending on the width of the excluded region, and this additional uncertainty
has been included in the adopted values shown in table~\ref{RVFitTable}.

%
%
\begin{figure*}
\caption{\label{EWFig} Variation of the equivalent width of the \halpha\
(filled circles) and Ca\,II\,8542\AA\ (crosses) emission lines with phase. Solid
lines are fits of the form $A + B\sin(\phi) + C \sin(2\phi)$. The dotted line
is the best fit for an optically thick model to the Ca\,II\,8542\AA\ line.}
\psfig{file=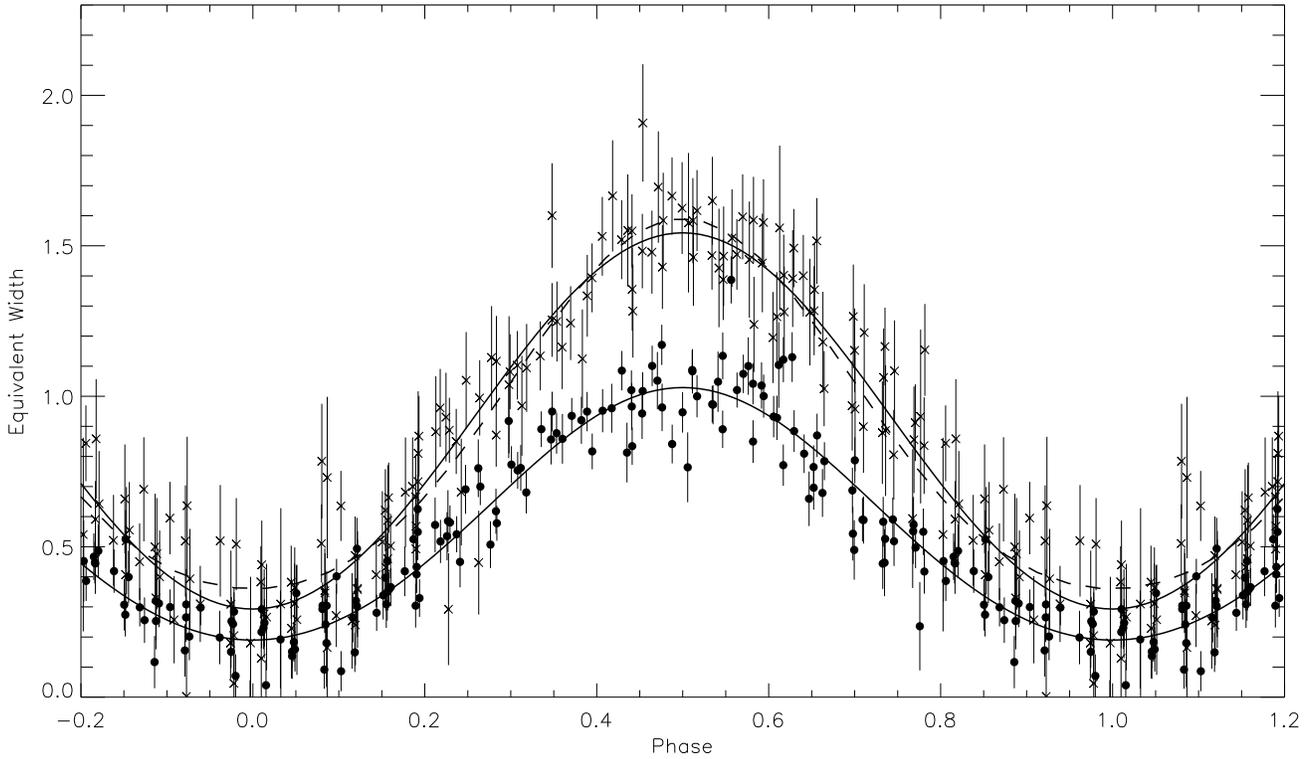}
\end{figure*}

 Our fit to the white dwarf absorption line was used to subtract the
contribution of the white dwarf from the \halpha\ line. The remaining M dwarf
emission line spectra were coadded in phase-bins to produce the trailed
spectra shown in Fig.~\ref{TrailFig}.  The equivalent width variation of this
line is shown in Fig.~\ref{EWFig}. It is obvious from Fig.~\ref{TrailFig} that
the \halpha\ emission is much wider than the Ca\,II emission lines, a feature
noted by MD96. Less obvious but still visible is the double-peaked shape of
the \halpha\ emission near phases 0.25 and 0.75. These anomalies are discussed
in more detail in section~\ref{DISCUSSSECTION}.

\subsection{\label{MassSection}The mass of the white dwarf and the M dwarf.}

%
%
\begin{table}
\caption{Summary of the parameters of GD\,448.
\label{ParamTable}}
\begin{tabular}{lrr}
Parameter & white dwarf & M dwarf \\
Mass (\Msolar)  &$0.41\pm0.01$&$ 0.096\pm0.004$ \\
Radius(\Rsolar) &$0.018          $& 0.125\,$\pm$\,0.02\,\Rsolar\\
Inclination(\degs) & \multicolumn{2}{c}{$29.3\pm0.7$} \\
Period  ($d$)    & \multicolumn{2}{c}{0.10306437} \\
\end{tabular}
\end{table}
 
 The measured values of  $K_{\rm WD}$, $K_{\rm M}$ and the orbital period lead
directly via Kepler's laws to $M_{\rm WD} \sin^3i$ and $M_{\rm M} \sin^3i$,
where $M_{\rm WD}$ is the mass of the white dwarf,   $M_{\rm M}$ is the mass of
the M dwarf and $i$ is the inclination. The actual mass of the white dwarf can
be established from the gravitational red shift and the mass\,--\,radius
relation for white dwarfs. To the  difference $(\gamma_{\rm WD}-\gamma_{\rm
M})=13.9\pm0.6\kms$ measured from the \halpha\ line must be added corrections
for the red shift of the M dwarf (0.5\kms) and the difference in transverse
Doppler shifts (0.1\kms) as described by MD96, as well as for the potential at
the M dwarf due to the white dwarf (0.35\kms) and {\it vice versa}
(-0.08\kms), a correction which MD96 omitted. The mass of the white dwarf is
then determined from the model cooling curves for low-mass helium white dwarfs
of Althaus \& Benvenuto (1997). For white dwarfs with an effective temperature
of 19\,000K and a mass near 0.4\Msolar, we find the mass in solar masses is
related to the gravitational red shift, $v_g$, in \kms\ by  $M=0.181+0.0157
v_g$. The inclination can then be determined using equation (1) of MD96.  The
corrections to the gravitational red shift depend slightly on the assumed
values of $i$ and so an iterative scheme is used to find consistent values for
all the parameters. The dependence of the results on the assumed radius of the
M dwarf is negligible for any reasonable estimate. The inclination and masses
derived are given in table~\ref{ParamTable}. The radius of the white dwarf is
also given, although no error is quoted as this is strongly correlated with
the uncertainty in the mass.

\section{Determination of the radius of the M dwarf from the Ca\,II triplet.}
 
 The Ca\,II emission in GD448 is due to emission from the heated face of the
M dwarf so $K^{\rm Ca\,II}_{\rm M}$ is lower than $K_{\rm M}$. The correction
$\Delta K = K_{\rm M}-K^{\rm Ca\,II}_{\rm M}$ depends on the pattern of the
emission over the surface of the M dwarf and its radiative properties (e.g.,
optical depth, limb darkening) as well as the geometry of the system. The only
free parameter in the system geometry is the filling factor of the
M dwarf, $f$, the ratio of the radii of the M dwarf and its Roche lobe
measured from the centre-of-mass to the inner Lagrangian point. A value of
$f=1$ corresponds to a Roche lobe filling star. The
other parameters (e.g., inclination, mass ratio) have been derived in
section~\ref{MassSection} above. 

 For our model of the emission from the M dwarf, we assumed a source function
of the form $1 + \epsilon\tau\mu$ where $\tau$ is the optical depth along the
line of sight and $\mu$ is the cosine of the angle between our line of sight
and the normal to the surface. The dependence of the emission line flux from a
point on the star, $F$, with viewing angle then has the form: \[ F \propto
(1+\epsilon\cdot\mu)(1-e^{-\tau_0/\mu})-\epsilon\cdot\tau_0 e^{-\tau_0/\mu},\]
where $\tau_0$ is the vertical optical depth through the emitting region,
which is assumed to be constant over the surface of the star. In the optically
thick case ($\tau_0 \gg 1$), $\epsilon$ is related to the standard linear limb
darkening parameter which is given by $\epsilon/(1+\epsilon)$ when $\epsilon$
is positive. In cases where $\epsilon$ is negative, i.e., source function
increasing with height, as might be expected in the case of an irradiated
atmosphere, $\tau_0$ must be less than $-1/\epsilon$ to avoid negative source
function values.

 We used numerical integration over a model star defined by a surface of
constant Roche potential to predict emission line strengths and radial
velocities for various values of $f$, $\epsilon$ and $\tau_0$ assuming the
emission line strength to be proportional to the incident flux per unit area
from the white dwarf.  These were calculated at the same phases as the
observed data and given the same weighting in a sine fit that was used to
calculate $\Delta K$. This has an almost linear dependence on $f$ and so it is
very easy to find $f$ which gives the correct value of $\Delta K$.

A further constraint on our model of the Ca\,II emission comes from the
lightcurve of the Ca\,II lines, i.e., the variation of equivalent width with
phase.  By fitting a function of the form  $A + B\sin(\phi) + C
\sin(2\phi)$  to the observed equivalent width variation, we find $C/B
= -0.02 \pm 0.02$. This ratio describes the magnitude of any non-sinusoidal
component   of the lightcurve (i.e., its shape) independent of any zero-point
or scaling applied. For each value of $\epsilon$ and $\tau_0$ used, the
predicted equivalent widths for the optimum filling factor $f_0$ were
calculated at the same phases as the observed data and given the same
weighting in a fit of the same function. The ratio $C/B$ from these fits rules
out models with values of $\tau_0 \goa 1.0$. These models show a non-sinusoidal
component that is much larger than that observed ($C/B \sim 0.1$), i.e., they
are the wrong shape.  This is shown in Fig.~\ref{EWFig}.

%
%
\begin{figure}
\caption{\label{RocheFig}
A cross section through the orbital plane of GD\,448 showing the relative
sizes of the white dwarf (filled circle) and  the M dwarf (partially hatched)
with its Roche lobe (dashed lines). The hatched region of the M dwarf shows
approximately where the Ca\,II and \halpha\ arise. The $+$ symbol marks the
centre of mass of the system and the $\times$ symbol the centre of mass of the
M dwarf. The axes are in units of the orbital separation
($r_x$ and $r_y$) and the amplitude of the sine and cosine components 
of projected radial velocity ($K_x$ and $K_y$). } 
\psfig{file=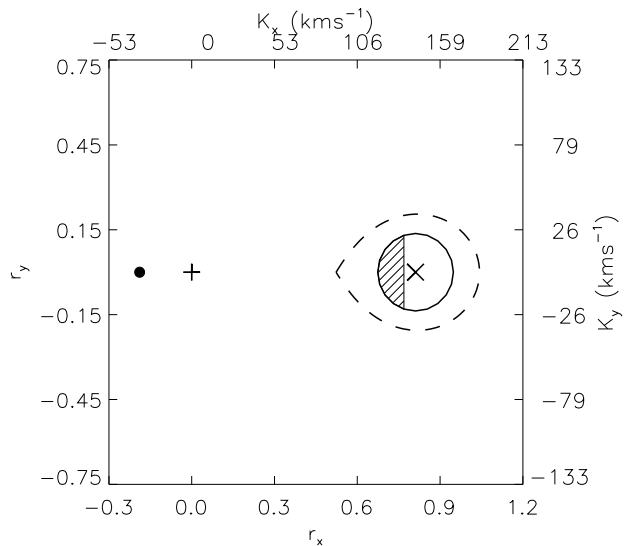} 
\end{figure}

 The values of $f_0$ found for valid values of $\tau_0$ vary from 0.46
($\tau_0 \ll 1$) to 0.49 ($\tau_0=1$) whatever the value of $\epsilon$. The
uncertainty in $K_{\rm M}$ leads to an further uncertainty of 0.06 in $f_0$.
The absolute radius of the M dwarf is then
0.125\,$\pm$\,0.015\,$\pm$\,0.005\,\Rsolar, where the uncertainties quoted are
the random error inherited from $K_{\rm M}$ and the systematic error due to
the range of $f_0$ values, respectively. The combination of systematic and
random uncertainties is a moot point, but the result cannot be worse than a
simple linear addition and so we will adopt a value of
0.125\,$\pm$\,0.02\,\Rsolar. The relative sizes of the white dwarf, the M dwarf
and the Roche lobe of the M dwarf are shown in Fig.~\ref{RocheFig}.

%
%
\begin{figure}
\caption{\label{MvRFig}
GD\,448 in the mass-radius plane (filled circle) compared to other low-mass
stars and brown dwarfs from Clemens et~al. (open circles).}
\psfig{file=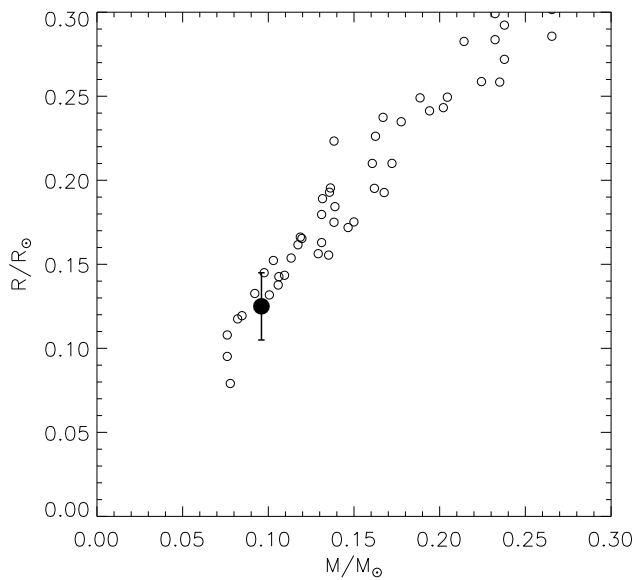}
\end{figure}

 The position of the M dwarf in the mass--radius plane is shown in
Fig.~\ref{MvRFig}. Also shown in Fig.~\ref{MvRFig} for comparison  are masses
and radii for other M dwarfs given by Clemens et~al. (1997). Their masses  are
based on their revised  empirical color--luminosity relation and the
luminosity--mass relation of Henry \& McCarthy(1993) based on 37 visual
binaries. The radii are derived from their updated colour--temperature and
colour--bolometric correction relations based on a volume limited sample of
127 M dwarfs. The agreement between their indirect method and our more direct
method is quite satisifactory.

\section{\label{DISCUSSSECTION} Discussion.}

 Comparing the position of GD\,448B in the mass--radius plane to those of other
M dwarfs, we see that it appears to be a perfectly normal main-sequence
M dwarf. The thermal timescale for the M dwarf ($3\times10^9$y) is much longer
than the cooling age of the white dwarf ($5\times10^7$y), so the radius will
not have changed significantly since the common-envelope phase. The argument
of MD96 which suggests that GD\,448 was born in the period gap, i.e., has never
been a CV, is not affected by anything presented here.

 The anomalous width of the \halpha\ emission line is particularly strange
given that its intensity variation with phase and radial velocity
semi-amplitude are almost identical to the much narrower Ca\,II emission lines.
For example, from a fit of the form $A + B\sin(\phi) + C \sin(2\phi)$  to the
equivalent width variation of the Ca\,II\,8542\AA\ line shown in
Fig.~\ref{EWFig} we find,  $B/A = -0.68\pm0.01$ and $C/B = -0.03\pm0.02$. The
other Ca\,II lines give similar results. For the \halpha\ emission we find the
remarkably similar values $B/A = -0.71\pm0.02$ and $C/B = -0.05\pm0.02$, i.e.,
the Ca\,II and \halpha\ emission lines have nearly identical amplitudes and
shapes. This appears to rule out any broadening mechanism requiring a
distribution of \halpha\ emitting material much different from the Ca\,II
emitting region, e.g., photospheric emission. 

Our favoured explanation for the shape of the \halpha\ line is some broadening
mechanism (e.g., natural damping or Stark broadening) within an optically thick
medium. This is at variance with the results of our simple model for the
emission of Ca\,II which appears to rule out emission from an optically thick
gas. However, our simple model is not complete as it does not account for any
variation of the optical depth over the surface of the M dwarf, nor for the
finite geometrical depth of the emitting layer. In the case of emission from
an optically thin gas, the emission line strengths are expected to be
proportional to the oscillator strengths of the transitions. In the case of
the Ca\,II lines, these ratios are {\mbox
Ca\,II\,8498\,:\,Ca\,II\,8542\,:\,Ca\,II\,8662 = 1\,:\,9\,:\,5}. The observed ratio
of the emission line strengths is 1\,:\,1.4\,:\,1.2, which suggest that at
least some part of that emission is due to optically thick material. The
double-peaked shape of the \halpha\ emission might then be explained as
self-absorption of the line. Clearly, more detailed modeling would be required
to demonstrate that these suggestions are feasible, but if they do adequately
explain the shape of the \halpha\ and Ca\,II emission lines, these lines would
then be giving information on the density and temperature structure within the
emitting region. 

 If our assumption that the pattern of Ca\,II emission is proportional to the
irradiating flux is wrong, it might seem possible to explain the width of
\halpha\ as rotational broadening of an intrinsically narrow line. This would
then require a very odd distribution of \halpha\ and Ca\,II emission over the
surface M dwarf to explain the shape of the \halpha\ line.
In fact, we used a very general model of the \halpha\ emission in which the
pattern of emission over the surface of the M dwarf was allowed to vary
independently at every grid point, each of which emits a narrow line. The
model includes the effects of orbital phase smearing and the instrumental
resolution. We then attempted to use a maximum entropy reconstruction to
determine a pattern of emission that produced emission lines consistent with
the data, but were unable to find any such pattern for any value of $f$. This
is a similar problem to that found by Rutten \& Dhillon (1994) in the case of
the cataclysmic variable star DW~UMa, although that case is complicated by the
presence of an accretion disc.

 The photometry for GD448 has only been used to confirm the correct orbital
period for the system and to normalise the continuum value for the red-arm
spectra. A simple model of the irradiation shows that the amplitude of the
lightcurve is consistent with that which would  be expected from an irradiated
star, but given the uncertainties in the heating mechanism and the small
amplitude of the modulation this  gives no constraint on the system geometry
in practice.

 The problem of the strength of the \halpha\ emission discussed by MD96
remains and is made worse by the small radius for the M dwarf derived here.
Briefly, there is insufficient radiation below the Lyman limit from the
white dwarf incident at the M dwarf surface to produce the observed emission
line strength by photoionisation of hydrogen. A similar problem has been
identified for the detached white dwarf\,--\,M dwarf binaries PG1026+002
(Saffer et~al., 1993) and WD2256+249 (Schmidt et~al., 1995). MD96 suggest that
chromospheric activity may excite hydrogen atoms to the n=2 level, from where
they are ionised. 

 An alternative explantion comes from the possibility of accretion onto the
white dwarf from the M dwarf wind, a process seen in the white dwarf\,--\,K
dwarf detached binary V471~Tau (Mullan et~al. 1991). The resulting irradiation
from the soft X-rays produced would be sufficient to produce the observed
\halpha\ flux for mass-loss rates only a few times greater than the Solar
mass-loss rate. To calculate the \halpha\ flux, we first used the
mass\,--\,colour\,--\,absolute-magnitude  relations for M dwarfs from Clemens
et~al. (1998) to find (V-I)=$4.0\pm0.1$ and $\Mv=16.2\pm0.3$. We then used
\Mv=10.15 for the white dwarf (Bergeron et~al., 1992)  and (V-I)=$-0.204$ for
a hydrogen-rich white dwarf of 19\,000K  (Bergeron et~al.1995) to predict a
contribution of 15\%$\pm$3\% from the M dwarf in the I band ($\sim 8200\AA$).
This is in perfect  agreement with the estimate of MD96. We used (R-I)=2.16
for the M dwarf (Bessel, 1991), which implies a spectral type of M5.5\,--\,M6,
and (R-I)=-0.11 for the white dwarf (Bergeron et~al. 1995) to find a
contribution of 2.2\%$\pm$0.7\% from the M dwarf in the R band ($\sim
7000$\AA). The peak equivalent width of the \halpha\ emission relative to the
combined continuum is 1.0\AA\ so the true equivalent width of the emission is
$\sim 45$\AA. The zero-point of the R-band magnitude scale for M dwarfs is
approximately $30\times 10^{-13}$ W\,m$^{-2}{\rm \AA}^{-1}$ (Allard \&
Hauschildt. 1995) and the V magnitude of GD\,448 is 14.97 (Bergeron et~al.,
1992) so the observed \halpha\ flux from the M dwarf is $\approx
3\times10^{-18}$W\,m$^{-2}$. For the purposes of this discussion we can assume
this \halpha\ flux is radiated isotropically from the M dwarf, in which case
the luminosity in \halpha\ of the M dwarf is $3\times10^{20}$W. If all the
X-radiation emitted from the white dwarf intercepted by the M dwarf (1.5\% for
isotropic emission) is re-processed as \halpha\ emission, the X-ray luminosity
of the white dwarf is $2\times10^{22}$W. The mass accretion rate required to
provide this energy is $\sim 1\times10^{-14}\Msolar {\rm y}^{-1}$. The
efficiency of Bondi-Hoyle accretion from the M dwarf wind is expected to be
quite high ($\sim 20$\%, Mullan et~al. 1991). Furthermore, models of X-ray
heating of dwarf M-star chromospheres suggest that perhaps half of the energy
is re-emitted as Balmer emission if the X-rays are absorbed high in the
atmosphere (Cram 1982). The total mass-loss rate required to produce \halpha\
could then be as low as $\sim 1\times10^{-13}\Msolar {\rm y}^{-1}$  --
comparable to the Solar mass-loss rate and much lower than the observed
mass-loss rate in other M dwarfs (Mullan et~al. 1992). The observed X-ray flux
in this scenario would be $\sim 2\times10^{-16}{\rm W\,m}^{-2} \equiv
2\times10^{-13} {\rm ergs\,s}^{-1}{\rm cm}^{-2}$, which is well within the
capabilities of the current generation of X-ray satellites.

\section{Conclusion.}

 Extensive analysis of spectra covering the \halpha\ line, the Ca\,II triplet
and the NaI doublet from GD448 have enabled us to determine the mass and
radius of the M dwarf companion to the white dwarf GD\,448. Its position in
the mass-radius relation places it squarely on the main-sequence. This
suggests that the common envelope phase had little effect on the structure of
the M dwarf. The anomalous width of the \halpha\ emission from
the M dwarf remains to be explained. The anomalous strength of the \halpha\
line may be due to X-ray heating of the M dwarf due to accretion onto the
white dwarf from the M dwarf wind.

~
\label{lastpage}

\end{document}